\documentclass[aps,prl,twocolumn,showpacs,superscriptaddress]{revtex4}
\begin{document}
\title{Cosmological Parameters are Dressed}
\author{Thomas Buchert}\thanks{buchert@theorie.physik.uni-muenchen.de}
\affiliation{Theoretische Physik, Ludwig--Maximilians--Universit\"at,
Theresienstr. 37, D--80333 M\"unchen, Germany \\
and Department of Physics and Research Center for the Early Universe
(RESCEU), School of Science, The University of Tokyo, Tokyo 113--0033, Japan}
\author{Mauro Carfora}\thanks{mauro.carfora@pv.infn.it}
\affiliation{Dipartimento di Fisica Nucleare e Teorica,
Universit\`{a} degli Studi di Pavia and 
Istituto Nazionale di Fisica Nucleare, Sezione di Pavia,
via A. Bassi 6, I--27100 Pavia, Italy}
\pacs{04.20 , 98.80 , 02.40 K}
\begin{abstract}
In the context of the averaging problem in relativistic cosmology, we provide 
a key to the interpretation of cosmological parameters by taking into account 
the actual inhomogeneous geometry of the Universe. We discuss the relation 
between `bare' cosmological parameters determining the cosmological model, 
and the parameters interpreted by observers with a ``Friedmannian bias'', 
which are `dressed' by the smoothed--out geometrical inhomogeneities of the 
surveyed spatial region.
\end{abstract}
\maketitle
%
%
A considerable body of researchers in cosmology holds the attitude
that, in view of ongoing and near future high--precision experiments in
observational cosmology, the (to a large extent established) 
standard model of cosmology will be finally 
determined. In particular, this attitude is supported by the possibility of precisely 
and unambigously determining the three numbers that fix the observer's position in the 
cosmic triangle and characterize the parameters of a FLRW cosmology \cite{bahcall}. 
Refinement of observational data, however, must
be paraphrased by a refinement of the theoretical cosmological model.
In particular, the widely hold working assumption that the standard
model idealizes in a dynamically consistent way
the real inhomogeneous Universe should be tested and questioned. For
instance, the conjecture that the standard model is equivalent to an
averaged inhomogeneous model cannot be held 
(e.g., \cite{ellis:relativistic}). Indeed, in this context, the
deviations of the average model from the standard model are 
condensed into a `backreaction effect'. It can be explored to understand the influence 
of structure inhomogeneities on the evolution of the standard model
parameters regionally, but may not impair the robustness of the conceivably simplest 
cosmological model on the largest scales. 

Complementary to this `backreaction effect', we, in this {\sl Letter}, want to
elaborate on a key--insight into the {\sl interpretation} of
cosmological parameters that, so far, escaped the attention of researchers 
in cosmology. It adds a new aspect to the discussion of the effect of inhomogeneites
on the standard model parameters, but -- moreover -- it provides an answer to the fundamental
problem of interpreting cosmological parameters in an inhomogeneous spacetime geometry.

Let us develop a picture that may guide our thoughts. 
Imagine a finite amount of material mass distributed inhomogeneously
within some spatial domain. We simplify the problem by assuming that the 
astronomical experiment is carried out in a sufficiently shallow survey region,
so that the observed objects, within the approximation standards we want to imply,
all lie in a `spatial' section and, 
referring to the space section itself, we assume that the theoretical model already 
gives us a suitable split into space and time, i.e., a foliation of spacetime. 
Suppose now that the observer would be able to quantify the observed objects by their
amount of material mass, employing, of course, some theoretical considerations, 
so that the simplest quantity that the astronomical experiment
returns is the total amount of material mass contained within the observed portion
of the Universe. This in turn determines, up to the normalization by the 
``Hubble--constant'' to which we come later, one of the standard cosmological parameters
on the scale of the observed region,
the {\sl density parameter}, if the amount of mass is divided by the surveyed volume.
It is here, where the `interpretation problem' comes into the fore: the ``observer's Universe'', 
due to a lack of better standards, is a constant curvature space section given by the standard 
model. Calculating the average density with the ``Friedmannian volume'' is, in
this picture, considered as the actual source in Friedmann's equation.

One of the reasons for this commonly held view is that Newtonian cosmology is 
the familiar framework of structure formation models, 
and the standard (constant) curvature 
parameter is merely taken to determine the `background' FLRW model, while structures
are described within a Euclidean homogeneous space geometry. 
The careful reader would object that the actual surveyed volume of the spatial 
domain is not the volume of a constant curvature FLRW domain, but -- taking the 
curvature fluctuations due to the inhomogeneities into account -- is rather the volume
of the bumpy geometry of the surveyed region. There is an obvious difference 
between the `bare' density parameter (the actual material mass density source), 
and the parameter obtained with a ``Friedmannian bias''. 
We may say that the latter is `dressed' by the geometrical inhomogeneities,
which the ``interpreter'' imagines to be smoothed--out, so that the averaged material mass density 
field is actually considered as an average over a homogeneous geometry. 
We are going to focus on the relation between `bare' and `dressed' parameters.
\section{Effective Cosmological Models}
To begin with let us recall a central result in connection with the averaging problem. 
It has recently been shown that a set of `generalized Friedmann equations',
which also incorporate structure inhomogeneities, govern the {\sl effective}
cosmological evolution \cite{buchert:grgdust}. `Effective' means that the 
homogeneous--isotropic variables of the FLRW model are replaced by their
Riemannian volume averages on some given spatial domain. In relativistic cosmology the 
generalized Friedmann equations, restricted here to the simplest matter model 
`irrotational dust' (more general matter models are discussed in \cite{buchert:grgfluid}), 
read:
\begin{eqnarray}
6 H^2_{{\cal B}_{0}} - 16\pi G
\langle\varrho\rangle_{{\cal B}_{0}} - 2 \Lambda + \langle{\cal
R}\rangle_{{\cal B}_{0}} = -  {\cal Q}^K_{{\cal B}_{0}}\;\;;
\label{friedmannregional1} \\
V_{{\cal B}_0}^{-2/3}\frac{d}{dt}\left(\langle {\cal R} \rangle_{{\cal B}_0} 
V_{{\cal B}_0}^{2/3}\right) \;+\;
V_{{\cal B}_0}^{-2} \;\frac{d}{dt}\left({\cal Q}^K_{{\cal B}_{0}} V_{{\cal B}_0}^{2}\right) \;=\;0
\;\;,
\label{friedmannregional2} 
\end{eqnarray}
where we have defined, on the averaging domain ${\cal B}_0$,
the {\em regional Hubble parameter} as $1/3$ of the spatially averaged
rate of expansion $\theta$:
\begin{equation}
3 H_{{\cal B}_0}: = \langle \theta \rangle_{{\cal B}_0}\;=\;\frac{1}{V_{{\cal B}_0}}
\int_{{\cal B}_0} \theta \;d\mu_g 
= \frac{\frac{d}{dt}V_{{\cal B}_0}}{V_{{\cal B}_0}}\;\;.
\label{hubble}
\end{equation}
$V_{{\cal B}_0} = \int_{{\cal B}_0} d\mu_g$ is the volume of the domain of averaging,
$d\mu_g$ the Riemannian volume element associated with the $3-$metric $g_{ab}$ of the 
hypersurface, ${\cal R}$ is the intrinsic scalar curvature, $\langle\varrho\rangle_{{\cal B}_0}:
= M_{{\cal B}_0}  V_{{\cal B}_0}^{-1}$ is the average matter density,
where $ M_{{\cal B}_0} = const.$, 
and $\frac{d}{dt}$ denotes the time--derivative in a comoving frame.
(Note that the zero--subscript indicates that the averaging domain
has the original non--averaged geometry; we shall later also refer to the corresponding 
smoothed--out domain.)

The explicit source term ${\cal Q}^K_{{\cal B}_{0}}$, 
the {\em kinematical backreaction}, appearing in the above equations quantifies the
deviations of the average model from the standard FLRW model.
It is composed of two positive--definite fluctuation terms (see: \cite{buchert:grgdust}):
first, {\em shear fluctuations} that tend to mimic the presence of a 
(kinematical) dark matter component decelerating the expansion and, second, 
{\em expansion amplitude fluctuations} that tend to mimic
a time--dependent positive cosmological term, an accelerating component (`quintessence').  

Eq.~(\ref{friedmannregional2}) shows that the averaged scalar curvature is coupled to the
`backreaction' dynamically, which is not the case in the corresponding Newtonian set of
equations \cite{buchert:average}. For ${\cal Q}^K_{{\cal B}_0}=0$ the set of
generalized Friedmann equations is closed, and we have from Eq.~(\ref{friedmannregional2}) 
$\langle{\cal R}\rangle_{{\cal B}_0} \propto V_{{\cal B}_0}^{-2/3}$ in agreement with the 
evolution of the spatially constant curvature in the standard model.

Furthermore, in the general model, we may define 
{\em regional cosmological parameters} as the following scale--dependent 
functionals \cite{buchert:grgdust}:
\begin{eqnarray}
\label{standardparameters}
\Omega^M_{{\cal B}_0} : = \frac{8\pi G M_{{\cal B}_0}}{3 V_{{\cal B}_0}
H_{{\cal B}_0}^2 }\; ;\;
\Omega^{\Lambda}_{{\cal B}_0}:= \frac{\Lambda}{3 H_{{\cal B}_0}^2 }
\;;\;
\Omega^{R}_{{\cal B}_0}:= - \frac{\langle{\cal R}\rangle_{{\cal B}_0}}
{6 H_{{\cal B}_0}^2}\;,\\
\label{backreactionparameter}
{\rm and}\;,\;\;\;\;\;\;
\Omega^{{\cal Q}^K}_{{\cal B}_0} := - \frac{{\cal Q}^K_{{\cal B}_0}}
{6 H_{{\cal B}_0}^2}\;\;,\;\;\;\;\;\;\;\;\;\;\;\;\;\;\;\;\;
\end{eqnarray}
obeying by construction the relation:
\begin{equation}
\Omega^M_{{\cal B}_0} \;+\; \Omega^{\Lambda}_{{\cal B}_0}\;+\; 
\Omega^R_{{\cal B}_0}\;+\; \Omega^{{\cal Q}^K}_{{\cal B}_0} \;=\; 1 
\;\;.
\label{omegaconstraint}
\end{equation}
In contrast to the standard FLRW cosmological parameters there are 
four players. In the FLRW case there is by definition no kinematical  
backreaction, ${\cal Q}^K_{{\cal B}_0}=0$.
Hence, the `effective cosmology' can be determined by a 
scale--dependent and regional `cosmic quartet' 
\cite{buchert:onaverage} rather than by a global `cosmic triangle' \cite{bahcall}.

It is generally agreed that quantitative investigations of the (kinematical) backreaction effect 
point towards two results: first, the contribution of $Q^K_{{\cal B}_0}$ to the `cosmic quartet' is
quantitatively small in sufficiently large expanding domains of the Universe (it may 
contribute significantly on cluster scales and below \cite{abundance} and may be attributed
to cosmic variance on large scales); second, the {\sl dynamical} influence of a non--vanishing 
backreaction on the other (standard) cosmological parameters can -- nevertheless -- be large, 
in other words, the values of the standard parameters found on a given
hypersurface at an evolved time are, in general, 
not related to their initial values according to the 
FLRW model (for an investigation in Newtonian cosmology see: \cite{buchert:bks}). 
\section{Dressing Cosmological Parameters}
Eq.~(\ref{omegaconstraint}) forms the basis of a discussion of cosmological parameters
as they determine the theoretical model. They may not be, however, directly accessible
to observations. 
Unlike in Newtonian cosmology, where the corresponding equations have 
a similar form \cite{buchert:average}, it is not straightforward to compare the above relativistic
average model parameters to observational data.
The reason is that the volume--averages contain information on the 
actually present {\sl geometrical inhomogeneities} within the averaging domain.
In contrast, the ``observer's Universe'' is 
described in terms of a Euclidean or constant curvature model.   
Consequently, the interpretation of observations within the set of 
the standard model parameters, if extended by the backreaction parameter or not,
neglects the geometrical inhomogeneities that
(through the Riemannian volume--average) are hidden in the average
characteristics of the theoretical cosmology. In other words, an averaging procedure in 
relativistic cosmology is not complete unless we device a way to also average the 
geometrical inhomogeneities. Since geometrical fields are tensorial variables for which
possible strategies of averaging form the subject of considerable 
controversy in the relativistic literature, there is some ambiguity in how such an 
averaging could be implemented.

We have suggested an answer to this problem in \cite{klingon}, and here we wish to
exploit our results for comparing the original averaged model of a surveyed
region of the Universe with the geometrically smoothed--out model which
governs the interpretation of the observer's data. 
This turns out to be rather simple and physically clear,
so that we think that explaining the details of the smoothing procedure is not mandatory in this 
{\sl Letter}. It suffices to say that the idea of smoothing--out the geometrical 
inhomogeneities was implemented in \cite{klingon} (see also \cite{carfora:RG} for a preliminary 
attempt), by designing a smoothing flow on the basis of the geometrical scaling properties 
of the matter variables. Moreover, such a smoothing was implemented on a regional and Lagrangian
basis, i.e., the metric and the matter variables are smoothed on a geodesic
domain in such a way as to preserve its material content. Such requirements
characterize in a natural way a Ricci deformation flow for the metric \cite{klingon}.
It is perhaps interesting to note that such a flow is extensively studied in
the mathematical literature (see, e.g.: \cite{hamilton:ricciflow2}, \cite{carfora:deformation1},
\cite{carfora:deformation2}), where the Ricci flow plays a basic role in
mapping a bumpy $3-$geometry into a homogeneous geometry.

Let us highlight some results. 

According to \cite{klingon}, the picture discussed in the introduction strictly
depends on the ratio between two density profiles defined in the averaging
domain: one is naturally associated with the actual matter content of the
gravitational sources, whereas the other is the mass density corresponding
to the matter content in the given region, but now thought of as averaged
over a geometrically smoothed--out domain $\overline{\cal B}$ with homogeneous geometry:
\begin{equation}
\langle \varrho \rangle_{{\cal B}_0} = M_{{\cal B}_0} / V_{{\cal B}_0}\;\;\;;\;\;\;
\langle \varrho \rangle_{\overline{\cal B}} = M_{\overline{\cal B}} / V_{\overline{\cal B}}\;\;.
\label{averagedensities}
\end{equation}
Our assumption of the regional smoothing was such that the total masses are the same, so that
we infer from (\ref{averagedensities}) that the average density measured with a 
``Friedmannian bias'' is dressed by a {\sl volume effect} due to the difference between the 
volume of a smoothed region and the actual volume of the bumpy region:
\begin{equation}
\langle \varrho \rangle_{{\cal B}_0} =\langle \varrho \rangle_{\overline{\cal B}}
( V_{\overline{\cal B}} / V_{{\cal B}_0}) \;\;.
\label{densityrelation}
\end{equation}   
A further result that explicitly involves the geometrical smoothing flows is formed
by the relation between the 
constant regional curvature in the smoothed model (e.g., a FLRW domain) and the actual
regional average curvature in the theoretical cosmology:
\begin{equation}
\label{regionalcurvature}
\overline{\cal R}_{\overline{\cal B}} = \langle {\cal R} 
\rangle_{{\cal B}_0} 
( V_{\overline{\cal B}} / V_{{\cal B}_0})^{-2/3}
- {\cal Q}^R_{{\cal B}_0} \;\;, 
\end{equation}
where we have introduced a novel measure for the `backreaction' of geometrical
inhomogeneities capturing the deviations from the standard FLRW space section, 
the {\em regional curvature backreaction}:
$\;{\cal Q}^R_{{\cal B}_0}:= \int_0^{\infty} d\beta \; \frac{V_{{\cal B}_{\beta}}(\beta)}
{V_{\overline{\cal B}}} \cdot$

\noindent
$\;\;\left[ 
\frac{1}{3}\langle \left({\cal R}(\beta) - \langle {\cal R}(\beta)
\rangle_{{\cal B}_{\beta}}\right)^2 \rangle_{{\cal B}_{\beta}}
- 2 \langle{\tilde{\cal R}}^{ab}(\beta){\tilde{\cal R}}_{ab}(\beta) 
\rangle_{{\cal B}_{\beta}}\right]\;$,  
with $\tilde{\cal R}_{ab}: ={\cal R}_{ab}-\frac{1}{3}g_{ab}{\cal R}$ being the trace--free
part of the Ricci tensor ${\cal R}_{ab}$ in the hypersurface.
${\cal Q}^R_{{\cal B}_0}$, built from scalar
invariants of the intrinsic curvature,
appears to have a similar form as the `kinematical backreaction' term (that was built
from invariants of the extrinsic curvature). It features two positive--definite parts, 
the {\em scalar curvature amplitude fluctuations} and {\em fluctuations in metrical anisotropies}. 
Depending on which part dominates we obtain an under-- or overestimate of the actual 
averaged scalar curvature, respectively.  
$\beta$ parametrizes integral curves of the smoothing flow for the metric,
so that the expression above indeed refers to the explicit form of this flow. 
Notwithstanding, this term may be estimated by the actual curvature fluctuations, since the 
Ricci flow acts in a controllable way such that the maxima of the curvature inhomogeneities
are monotonically decreasing during the deformation.

From  Eq.~(\ref{regionalcurvature}) we can understand the 
physical content of geometrical averaging. It makes transparent that, in the smoothed model, 
the averaged scalar curvature is `dressed' both by the {\em volume effect} mentioned above, 
and by the {\em curvature backreaction effect} itself. The volume effect is expected
precisely in the form occurring in (\ref{regionalcurvature}), if we think of comparing two regions
of distinct volumes, but with the same matter content, in a constant
curvature space. Whereas the backreaction term encodes the deviation of
the averaged scalar curvature from a constant curvature model, e.g., a FLRW
space section.
\section{The Bare Quartet}
The results discussed above allow us to relate the actual parameters (\ref{standardparameters}) and
(\ref{backreactionparameter}) to the values of such
parameters obtained as regional averages on a homogeneous geometry by the smoothing procedure.
We have seen that a ``Friedmannian bias'' in modelling the real observed region of the
Universe with a smooth matter distribution evolving in a homogeneous and
isotropic geometry, inevitably `dresses' the matter density $
\left\langle \varrho\right\rangle_{\overline{\cal B}}$, the
Hubble parameter ${\overline{H}}_{\overline{\cal B}}$, and 
the scalar curvature ${\overline{\cal R}}_{\overline{\cal B}}$
with correction factors.
Correspondingly, an observer with a ``Friedmannian bias'' would interprete
his measurements in terms of the `dressed' cosmological parameters:
\begin{eqnarray}
\overline{\Omega}^M_{\overline{\cal B}}:=\frac{8\pi G M_{\overline{\cal B}}}{3 V_{\overline{\cal B}}
\overline{H}_{\overline{\cal B}}^2 }\;;\;\overline{\Omega}^{\Lambda}_{\overline{\cal B}}:= 
\frac{\Lambda}{3 \overline{H}_{\overline{\cal B}}^2 }\;;\;
\overline{\Omega}^{R}_{\overline{\cal B}}:= 
-\frac{\overline{\cal R}_{\overline{\cal B}}}
{6 \overline{H}_{\overline{\cal B}}^2}\;;\;\;\;\;\;\;\;\;\;\;\;\;\\
\overline{\Omega}^{{\cal Q}^K}_{\overline{\cal B}}:= 
-\frac{\overline{\cal Q}^K_{\overline{\cal B}}}
{6 \overline{H}_{\overline{\cal B}}^2}\;,\;
{\rm obeying}\;\;\;
\overline{\Omega}^M_{\overline{\cal B}} + 
\overline{\Omega}^{\Lambda}_{\overline{\cal B}}+ 
\overline{\Omega}^R_{\overline{\cal B}}+
\overline{\Omega}^{{\cal Q}^K}_{\overline{\cal B}} = 1 
\,.\nonumber
\label{omegaconstraintdressed}
\end{eqnarray}
The latter equation follows from our assumption that the 
smoothing procedure requires to respect the Hamiltonian constraint of Einstein's equations. 
Introducing the dimensionless parameters
\begin{equation}
\nu : = \frac{V_{\overline{\cal B}}}{V_{{\cal B}_0}}\;\;\;;\;\;\;
\alpha : = \frac{\overline{H}^2_{\overline{\cal B}}}{H^2_{{\cal B}_0}}\;\;;\;\;
\mu : = \frac{{\cal Q}^R_{{\cal B}_0}}{{\overline{\cal R}}_{\overline{\cal B}}}\;\;,
\end{equation}
we can formally study fractions of `bare' and `dressed' 
parameters (making sure that the denominators are non--zero, 
which is the case in generic situations):
\begin{eqnarray}
\label{cosmologicalfractions}
\frac{\Omega^M_{{\cal B}_0}}{\overline{\Omega}^M_{\overline{\cal B}}} \;=\;
\alpha \; \nu \;\;;\;\;\frac{\Omega^{\Lambda}_{{\cal B}_0}}
{\overline{\Omega}^{\Lambda}_{\overline{\cal B}}} \;=\;
\alpha \;\;;\;\;\;\;\;\;\;\;\;\;\;\;\;\;\;\; \\
\frac{\Omega^R_{{\cal B}_0}}{\overline{\Omega}^R_{\overline{\cal B}}} \;=\;
\alpha \frac{\langle {\cal R}\rangle_{{\cal B}_0}}
{\overline{\cal R}_{\overline{\cal B}}} = \alpha\;\nu^{2/3}\;(1+\mu ) \;\;;\;\;
\frac{\Omega^{{\cal Q}^K}_{{\cal B}_0}}
{\overline{\Omega}^{{\cal Q}^K}_{\overline{\cal B}}} \;=\;
\alpha \frac{{\cal Q}^K_{{\cal B}_0}}
{{\overline{\cal Q}}^K_{\overline{\cal B}}} \;\;.\nonumber
\end{eqnarray}
The above listed relations appear to provide a formal recipee for interpreting the
cosmological parameters. Let us illustrate them 
by considering mixed fractions of various cosmological parameters in order
to eliminate, say the fraction of the Hubble parameters $\alpha$, and 
conclude on the values of the others:
\begin{equation}
\frac{\Omega^M_{{\cal B}_0}}{\Omega^R_{{\cal B}_0}} \;=\;
\frac{{\overline{\Omega}}^{M}_{\overline{\cal B}}}
{{\overline{\Omega}}^R_{\overline{\cal B}}}\frac{\nu^{1/3}}{1+\mu}
\;\;\;;\;\;\;
\frac{\Omega^M_{{\cal B}_0}}{\Omega^{\Lambda}_{{\cal B}_0}} \;=\;
\frac{{\overline{\Omega}}^{M}_{\overline{\cal B}}}
{{\overline{\Omega}}^{\Lambda}_{\overline{\cal B}}}\;\nu
\;\;.
\end{equation}
Reflecting the contemporary view on the cosmological parameters, we may consider 
a region of the Universe on a sufficiently large scale
of the order of $1$ Gpc/h. The (possibly also `dressed') 
observations of the first doppler peak in the CMB fluctuations at the 
``Friedmannian scale'' $\approx 100$ Mpc/h favour an approximately 
vanishing average curvature $\overline{\cal R}_{\overline{\cal B}}\approx 0$.
Let us, for simplicity, approximate both the `bare' and `dressed' `kinematical backreaction'
parameters by zero. 
If, again for simplicity, we approximate also the `curvature backreaction' parameter by zero,  
$\mu \approx 0$ (in the sense that there are curvature fluctuations present, but the 
two positive--definite parts compensate each other),
we would have an approximately vanishing average curvature also in the actual cosmological model.
Then, the standard argument requires compensation of the observed
matter content (including dark baryonic and possibly dark nonbaryonic matter
components), obeying the commonly agreed upper bound 
${\overline{\Omega}}^M_{\overline{\cal B}} \le 0.3$ with a cosmological term
${\overline{\Omega}}^{\Lambda}_{\overline{\cal B}} \approx 0.7$.
For the `bare' parameters we then obtain $\Omega^M_{{\cal B}_0} /
\Omega^{\Lambda}_{{\cal B}_0} \approx  \frac{0.3}{0.7}\nu$, which 
yields the estimate:
\begin{equation}
\Omega^M_{{\cal B}_0} \approx \frac{0.3}{0.7}\nu / ( 1+ \frac{0.3}{0.7}\nu )
\;\;\;;\;\;\;\Omega^{\Lambda}_{{\cal B}_0}\approx 1 - \Omega^M_{{\cal B}_0}\;\;.
\end{equation}
This certainly oversimplified example shows that, instead of postulating 
the presence of a large cosmological term, the `bare' mass parameter could still acquire 
values close to one, if `undressed', and if the volume fraction $\nu$ is substantially 
greater than 1. The second relation in Eq.~(\ref{cosmologicalfractions}) then shows, that
the actual Hubble--parameter would be larger than the `dressed' one.

A quantitative estimate that gives us an idea of the order of magnitude of the {\em volume effect}
has been worked out by Hellaby \cite{hellaby:volumematching} comparing spherically symmetric
with FLRW solutions. He employs ``volume matching'' as proposed by Ellis and Stoeger 
\cite{ellisstoeger} which should, however, amount to a similar effect 
as a comparison of the models at equal mass.
He finds that the spatial averages of the density profiles as compared with 
the corresponding (fitted) FLRW parameters yield errors in the range $10 - 30$\% for 
realistically modelled clusters and voids.

It appears that the interpretation of relativistic cosmological parameters is far from trivial,
given that we did neither touch on the issue of averaging on the observer's 
light--cone in which case the discussed effect interacts with the time--evolution 
of the model (compare the detailed suggestion in \cite{ellisstoeger}), nor
did we study the smoothing itself in a dynamical setting. 
As the present discussion shows, a thorough investigation of volumes of realistic 
cosmological slices as the ``simplest'' quantity would considerably enhance our theoretical 
background to approach observational data. As in other fields like solid state physics, where
the study of {\sl surface roughening} is well--developed, cosmology could face the necessity
of understanding geometrical structure formation, as it was facing the necessity of understanding
structure formation on a homogeneous geometry.  
\begin{acknowledgments}
TB would like to thank Martin Kerscher and Yasushi Suto for constructive remarks.
He acknowledges hospitality by the University of Tokyo and financial 
support by the Research Center for the Early Universe (RESCEU, Tokyo),
COE Monkasho Grant. This work is
also partially supported by the Sonderforschungsbereich SFB 375 
`Astroparticle physics' by the German science foundation DFG, and  
by the Ministero dell' Universita' e della Ricerca 
Scientifica under the PRIN project `The Geometry of Integrable Systems'.
\end{acknowledgments}

\end{document}